\documentclass{article}
\usepackage[latin1]{inputenc}
\usepackage{amsfonts}
\usepackage{amsmath}
\usepackage{amssymb}
\usepackage{braket}
\usepackage{enumerate}
\usepackage [english]{babel}
\usepackage [autostyle, english = american]{csquotes}
\MakeOuterQuote{"}
\usepackage{longtable}
\usepackage{mathrsfs}
\usepackage{titlesec}
\titleformat{\section}[block]{\fontsize{11}{15}\bfseries\filcenter}{\thesection}{1em}{}
\usepackage[left=0.875in, right=0.875in, top=1.00in, bottom=1.25in]{geometry}
\renewcommand{\thesection}{\Roman{section}}
\author{
Horace P. Yuen\\
\hspace{10mm}\\
Department of Electrical Engineering and Computer Science\\
Department of Physics and Astronomy\\
Northwestern University, Evanston Il. 60208\\
yuen@eecs.northwestern.edu
}
\date{}
\title{\LARGE \bf Essential lack of security proof in quantum key distribution\thanks{This paper with a similar title is to be published in the
Proceedings of the SPIE Conference on Quantum-Physics-Based
Information Security held in Dresden, Germany, Sep 23-24, 2013.
This v2 corrects some types in v1.}}
\renewcommand\thesection{\arabic{section}}

\begin{document}
\linespread{1}
\maketitle
\linespread{1}
\section*{ABSTRACT}
All the currently available unconditional security proofs on
quantum key distribution, in particular for the BB84 protocol and
its variants including continuous-variable ones, are invalid or
incomplete at many points. In this paper we discuss some of the
main known problems, particularly those on operational security
guarantee and error correction. Most basic are the points that
there is no security parameter in such protocols and it is not the
case the generated key is perfect with probability $\geq
1-\epsilon$ under the trace distance criterion $d\leq\epsilon$,
which is widely claimed in the technical and popular literature.
The many serious security consequences of this error about the QKD
generated key would be explained, including practical ramification
on achievable security levels. It will be shown how the error
correction problem alone may already defy rigorous quantitative
analysis. Various other problems would be touched upon. It is
pointed out that rigorous security guarantee of much more
efficient quantum cryptosystems may be obtained by abandoning the
disturbance-information tradeoff principle and utilizing instead
the known KCQ (keyed communication in quantum noise) principle in
conjunction with a new DBM (decoy bits method) principle that will
be detailed elsewhere.

\section{INTRODUCTION}
QKD (quantum key distribution) [1] protocols of the BB84 varieties
involving disturbance information tradeoff for security have been
widely claimed and perceived to provide "perfect security" at
reasonable key generation rate. This is the case in numerous
popular expositions and technical cryptography books which use
exactly the words "perfect security" [2], in major technical QKD
review articles that claim perfect security except with a very
small probability [3], and in numerous technical papers on
security theory and experimental implementations with the words
"unconditional security" [4]. This has been continuing despite the
criticisms on the invalidity of these claims, both fundamentally
in theory and empirically in practice. See [5] for some
references. It is the purpose of this paper to present an accurate
and readily understandable presentation of some fundamental points
in this connection, for proper appreciation of the scope and limit
of QKD.

Cryptographic security occupies a very unusual status compared to
most other issues in science and engineering. It cannot be
established experimentally in sufficient generality, if only
because there are unlimited classes of specific attacks, in
addition to numerous other issues. One is justified in claiming
security of a cryptographic scheme only by proving it rigorously
for a specific well defined mathematical model. Rigor is
important, many cryptosystems once thought secure turned out not,
and many others such as AES appear to be secure already. If only
seed key is needed, plenty can be stored compactly in many
applications. The issue of why a mathematical model is applicable
to a concrete QKD system  gives rise to major problems not found
in ordinary cryptography, where the security mechanism is based on
purely mathematical relations in contrast to QKD which involves
quantum effects of very small signals. Here we would just focus on
the security claims about specific given models. There are
numerous invalid inferences in every major step of the offered
security proofs, see [5] for brief descriptions of some of them.
Many will be mentioned but only a few will be discussed in the
following.

In this paper we will concentrate on one most fundamental issue,
the security criterion and its adequacy for \textit{operational
security guarantee}. We will identify the serious ramifications of
a major error of interpreting the trace distance criterion still
widely perpetuated today. Cryptographic security is a serious
business and has to be validated by theory, which cannot be
assured without scrutiny of the arguments offered for security
proof. As a concluding implication of our presentation, it will be
indicated that given the apparently insurmountable security proof
problems facing BB84 type protocols, major modification of
existing QKD protocols appears necessary in which the
disturbance-tradeoff principle is abandoned in lieu of other more
powerful principles for valid security proof. One such possibility
will be indicated.

\section{COMPARISON OF QKD WITH MATHEMATICAL CRYPTOGRAPHY}
In QKD two users Alice and Bob try to establish a new fresh key
between them by a protocol that involves five major steps outlined
in section IV of [5]. For general security the protocol should be
secure against an adversary Eve who could launch any attack
consistent with the laws of physics, active as wells as passive,
both during protocol execution and during actual use of the
generated key $K$ in a cryptographic application. Protocol
execution is interactive and requires message authentication
between the users. Some sort of information-disturbance tradeoff
is utilized by the users which requires checking a portion of the
quantum signals received by Bob to try to make sure they have not
been disturbed much in order that the information the attacker Eve
can extract from her attack on the other signals can be bounded
below a tolerable threshold. In this paper we can just take the
quantum signals to be qubits modulated by the digital data bit
sequence $X$ chosen randomly by Alice (except for CV-QKD in
section 6).

QKD has always been compared to RSA or public key cryptography to
this day, emphasizing its information theoretic security (ITS) as
compared to the complexity based security of RSA. I pointed out in
[6] that QKD should be compared to symmetric-key expansion which
\textit{also} offers ITS for the key security before it is used
and the ITS level of such symmetric-key ciphers is quite
\textit{good compared} to QKD systems. A prior shared secret key
is also needed in QKD as in symmetric-key expansion, at least for
message authentication during protocol execution needed to thwart
man-in-the-middle attack. What is unsatisfactory about ordinary
symmetric-key expansion is that it has no ITS under
known-plaintext attack (KPA) when the expanded key is used.

How does a known-plaintext attack work? Consider the case where
$K$ is broken into two segments $K=K_1||K_2$ and used to OTP
(one-time pad), or more accurately in an additive stream cipher to
xor into the data $X'=X_1'||X_2'$ in an encryption application of
$K$. We use sequence concatenation $||$ here for simplicity
instead of subset disjoint union. In KPA let Eve know $K_1'$
exactly for simplicity. The cipher text $K\oplus X'$ is always
open. Then Eve knows $K_1$ exactly and can use it to derive
knowledge of $K_2$ to help identify $X_2'$ with the known
$K_2\oplus X_2'$. If $K$ is not uniformly distributed to Eve,
there may be correlation between $K_1$ and $K_2$  that strongly
compromises $X_2'$. The two significant points are that the
precise quantitative ITS level of $K$ is important and that for
encryption it is KPA security that the advantage of QKD consists
in.

It is evident that there can be no KPA on $X$ chosen by Alice.
However, when $K$ is used for OTP $X'$, KPA becomes possible for
commercial applications at least. If KPA is not possible in some
military applications so that $X'$ is uniform to Eve, $K$ is
\textit{totally protected} and one can just use the
ciphertext-only ITS from a running key generated by standard
pseudo-random number generators. Some indication on how much more
secure that is as compared to any realistic QKD key has been
provided in [6], and a completed demonstration will be given
elsewhere.

\section{MUTUAL INFORMATION, SECURITY CRITERION AND SECURITY PARAMETER}
The security criterion commonly employed in QKD for a long time,
including in all the well known QKD asymptotic security proofs and
many experimental claims, is Eve's mutual information on $K$ that
she could obtain from measurement on her quantum probe together
with the side information she collects during protocol execution
and during use of $K$ in an application. In this quantum case it
is often called her "accessible information," the maximum mutual
information she could obtain by any quantum measurement. There is
no need to consider specific quantum feature here, we can take her
quantum measurement to be an appropriately optimal one. Thus, in
obvious notation,
\begin{equation}
I_E\equiv H(K)-H(K|E)
\end{equation}
Let $n$ be the bit length of the generated key $K$, $|K|=n$,
before accounting for any key cost during protocol execution. They
asymptotic security proofs assert that below a fixed nonzero
threshold key rate $r$ when the protocol is not aborted from qubit
checking due to too much disturbance, one may obtain
\begin{equation}
I_E\to 0 \quad\:as\:\quad n\to\infty
\end{equation}
With (2), $n$ is taken to be a cryptographic \textit{security
parameter} which means security can be made arbitrarily close to
perfect by making the security parameter arbitrarily large. This
is what the "perfect security" claim was then based upon. More
precisely, a protocol having \textit{unconditional security} [4]
means there is a security parameter for arbitrary attacks
consistent with the laws of physics, and in fact the key length in
a QKD round of key generation is taken to be such a security
parameter.

It has been pointed out repeatedly since 2004 [7], through several
talks including [8] and a full description in [9], that it the
rate that $I_E$ goes to zero that determines the security of $K$,
not $I_E$ itself. This will be explained and illustrated below.
But first the proper description of Eve's \textit{information} of
$K$ from her attack should be given, as follows [9].

From her probe measurement result $y_E$ together with side
information, Eve derives a whole probability distribution
$p(k|y_E)$ obtained from $p(y_E|k)$ via Bayes rule, and $p(y_E|k)$
through $\rho^k_E$, her probe state [9]. We suppress $y_E$ and
write $p(k|y_E)$ as $P={P_i}$ on the possible values of $K$. Let
$N=2^n$, and order $P_1\geq ... \geq P_N$. One should compare $P$
to $U$, the uniform distribution of $N$ values. This is evidently
the case if one looks at the quantitative security problem as a
detection theory problem for correctly identifying various subsets
of $K$ given whatever information, such as that of KPA in
identifying $K_2$ (of the previous section) from knowing $K_1=k_1$
(lower case denotes specific value of the upper case random
variable). $M$-ary detection is the appropriate framework for
security analysis, not Shannon information, for several reasons.
Mutual information is a theoretical quantity, one has to provide
\textit{operational guarantee} on Eve's probabilities of success,
and bit error rate (BER) when she fails to identify a subsequence
of $K$ but nevertheless gets a lot more bits correctly compared to
the perfect BER 1/2 from a uniform key, similar to the case of a
non-uniform a priori distribution of $K$ known to Eve. It is clear
the quality of $K$ and its quantitative level must be compared to
that of the uniform $U$, and there is no reason to conclude a
priori that any theoretical quantity is an adequate criterion when
it is not zero (with zero corresponding to perfect security).
"Information" is just a technical term in this connection. One
cannot read too much into the statement "Eve's information is
small." How quantitatively small is small enough for what purpose?

Thus, generally one needs the whole $P$ Eve possesses to be able
to assess various security details, which appears to be an
impossible task to estimate usefully in a QKD protocol. One may
use a single-number criterion such as $I_E$ and try to assess what
operational guarantee may be derived from it. Note that a
single-number guarantee \textit{merely} expresses a constraint on
$P$. For security guarantee one must show none of the $P$ not
ruled out by the constraint would allow undesirable information
for Eve. In this connection, Eve's maximum probability $P_1$ in
identifying the whole $K$ is especially crucial. Its average over
the a priori distribution of $K$ is simply related by $-log$ to
the so called minimum entropy, $H_{min}$. Note also that there is
no useful sense to talk further about the probability of a
particular $P$. Eve chooses for whatever reason a specific unknown
attack to the users with a resulting $P$ that is only constrained
by the single-number criterion. It was shown [7-9] that under an
$I_E$ guarantee, $P_1$ can be relatively very large with some
possible $P$. From Lemma 2 in [9], for $l<n$ it is possible that
\begin{equation}
\frac{I_E}{n}\leq 2^{-l}\:,\: P_1\sim 2^{-l}
\end{equation}

Since $l$ is typically very much smaller than $n$ with or without
privacy amplification, the $P_1$ of (3) is very much larger than
that of a uniform $K$. Note that $P_1$ typically increases, and
surely cannot be decreased by any privacy amplification code (PAC)
which is a many-to-one deterministic transformation. It is clear
the $P_1$ is a main determining factor on the security of $K$. PAC
may increase overall security for the same or larger $P_1$ because
the key after PAC is shorter than before and hence the new $P_1$
is effectively less damaging or can even be close to ideal. From
(3) it follows that,  $\frac{I_E}{n}\leq 2^{-l}$ strictly limits
the amount of (near) uniform key bits that can be obtained from
$K$ by further PAC, and strictly limits the security level of such
$K$ by itself. This $P_1$ consideration applies to the trace
distance criterion similarly, as is the operational meaning
problem.

             It follows from (3) that a very poor insecure $K$ can have $I_E\to 0$ for $n\to\infty$. Let
\begin{equation}
I_E=2^{-(\lambda n-\log{n})}
\end{equation}
for a constant $\lambda$. Then $I_E$ goes to  zero exponentially
in $n$ but it is possible that $P_1$ is given by $2^{-\lambda n}$
compared to $2^{-n}$ of a uniform key. It is clear that however
long $n$ is, if $\lambda<<1$ the key $K$ is then always very poor
compared to $U$ and it never improves relatively for any $n$. As a
consequence, $n$ is \textit{not} a security parameter at all in
QKD. There is in fact \textit{no} security parameter in any known
QKD protocol. Thus there can be \textit{no} unconditional security
proof for QKD in the original sense of the term [4].

             In this connection, it may be observed that asymptotic results are not sufficient for quantifying the performance of a realistic system in any event. Everything in the real world is finite with no parameter value approaching infinity. It is the quantitative security behavior of real systems that one must be concerned with. In particular, the so-called \textit{secrecy capacity} has no real security significance for concrete cryptosystems for two major reasons beyond finite versus asymptotic. The secrecy capacity is defined as the difference between the information capacities of the users and Eve. Since Eve cannot do coding on the data her capacity overestimates her information gain in general. Since the users don't know what active attack Eve has launched, there is no guarantee they can achieve their information capacity apart from the finite $n$ issue due to the lack of attacker channel characterization. Thus, the difference in capacities or mutual information between them has \textit{no} definite meaning. Note also that such capacity is derived from a constant channel among different uses, which simply does not obtain under active attacks and especially under entanglement attack in QKD. As indicated above, mutual information is not the proper framework for analyzing QKD security, $M$-ary detection is more appropriate.

                It turns out the phenomenon of quantum information locking makes $I_E$ not a good security criterion [10,11]. In [12] it is claimed that the system is secure if $I_E$ is exponentially small. Such claim is vague as can be seen from (4) above unless $\lambda$ is specified. It is falsified from [11] generally, since it is not ruled out that knowing $\log{n}$ data bits in a KPA on $K$ would reveal the entire $n$-bit $K$. Thus, the trace distance criterion $d$ is suggested instead [10, 13-14] while $d$ is first
proposed in [12] for composition security. The latter is a vague
concept because no operational quantitative meaning is ever given
for $d$, nor is any precise definition provided for "universal
composition". These problems are automatically solved, however,
under the prevalent incorrect interpretation of $d$.

\section{THE TRACE DISTANCE CRITERION}
                    The criterion $d$ is the trace distance between the real and the ideal quantum states for the users [10-14] and can be written as a $K$-average distance [15], with $p_0(k)$ being the prior probability of $K$,  $\rho^k_E$   Eve's probe state for each $k$, and $\rho_E$ the $k$-average $\rho^k_E$,
                    \begin{equation}
                    d=\frac{1}{2}\sum_k||p_0(k)\rho^k_E-\frac{1}{N}\rho_E||_1
                    \end{equation}

Generally $0\leq d\leq 1$, the smaller $d$ is the more secure $K$
is with $d=0$ for the perfect case. Upon measurement on Eve's
probe, trace distance becomes or bounds a classical statistical
distance $\delta(P,Q)$ between two classical distributions $P$ and
$Q$. The interpretation of an "$\epsilon$-secure" key, namely $K$
with $d\leq\epsilon$, amounts to saying perfect security is
obtained with a probability $\geq 1-\epsilon$ or equivalently the
"maximum failure probability" is $\epsilon$. For some specific
quotes see note [25] in [6]. We will call such interpretation of a
$d\leq\epsilon$ key an "$\epsilon$-uniform" key.

This error is maintained in the review [3] and has never been
re-tracked in the literature, contributing to the widespread
misconception mentioned in the Introduction. It has enormous
consequences on the quality of a QKD generated $K$, which becomes
far inferior to a uniform key. Before discussing such consequences
we summarize three possible reasons for holding such a wrong
interpretation and why they cannot be valid. Indeed, it can be
said that for $d>0$ the key $K$ is \textit{not uniform} with
probability 1 in general rather than uniform with probability
$\geq 1-\epsilon$. Actually, the mere talk of such probability is
already misleading because there is no general meaning one can
give to such probability.

                                The statistical distance $\delta(P,Q)$ defined by
                                \begin{equation}
                                \delta(P,Q)\equiv\frac{1}{2}\sum_i|P_i-Q_i|
                                \end{equation}
is interpreted in [14, 15] as the probability that $P$ and $Q$ are
the same except with a probability at most $\delta(P,Q)$. This is
obtained through a joint probability that gives $P$ and $Q$ as
marginals from a mathematical lemma that does not guarantee its
existence. Such a joint probability does not make sense since
there is no random source giving rise to it, and it does not imply
the interpretation even if it is in force. See [6, 9, 15] for
further discussion of this first reason. The second possible
reason is the "distinguishability" probability interpretation of
$d$ or $\delta$, which neglects that there is an additive factor
of 1/2 for binary decision [16]. The third possible reason is the
following decomposition under $\delta(P,U)=\delta_E$,
\begin{equation}
P_i=(1-\lambda)U_i+\lambda P_i'
\end{equation}
for another distribution $P'$ on $i\in\overline{1-N}$

We would not go into why (7) does not fully imply the wrong
interpretation when $\lambda=\delta_E$. It can be readily shown
that (7) holds if and only if
\begin{equation}
\frac{1-\lambda}{N}\leq P_i\leq\lambda + \frac{1-\lambda}{N}
\end{equation}
which implies all $P_i$ are essential uniform. In contrast, under
$d\leq\epsilon$ it is possible to have [9,15]
\begin{equation}
P_1=\frac{1}{N}+\epsilon
\end{equation}
Note that $\epsilon=10^{-20}$ [20] is very small for a binary
decision problem but very big for an $N$-ary decision problem for
$N=2^{1,000}$ and much larger in QKD protocols. In addition, there
is the problem of many repeated uses.

The failure and consequence of the wrong interpretation of $d$ can
be easily seen from the following result. Under
$\delta(P,Q)\leq\epsilon$, it is well known [17] that for an
unconditional event $A$,
\begin{equation}
|P(A)-Q(A)|\leq\epsilon
\end{equation}
For $Q=U$, it is the case [18] that a conditional event $B$ given
$A$ may achieve the following bound for some $P$ under the given
constraint,
\begin{equation}
|P(B|A)-U(B|A)|\leq\frac{\epsilon}{U(A)}
\end{equation}
The rhs of (11) can be much larger than $\epsilon$ and in fact
exceed 1, which means $P(B|A)$ is not constrained and can reach 1.
Under the wrong interpretation, on the other hand, $P(A|B)=
U(A|B)=\frac{|A|}{|B|}$ with  probability $1-\epsilon$. The
enormous difference in KPA security implication is obvious.

                                   Such possible drastic breach of security may be expected to average out over different possible conditioning. Indeed it is shown in [15] that one recovers, for the $K_1$ and $K_2$ averaged $\overline{P_1}(K_2^*|K_1$ with $K=K_1||K_2$ and $K_2^*$ any subset of $K_2$ that, for $\delta_E\leq\epsilon$,
 \begin{equation}
 \overline{P_1}(K_2^*|K_1)\leq 2^{-|K_2^*|}+\epsilon
\end{equation}
Note that not just the whole $K$, not subset $K_2^*$ has a better
protection that $\epsilon$ from (12). Contrast such $K$ average
with the  case of$ $U which holds for any $k$ and $k$-subset. Due
to such additional averaging, instead of just averaging over PAC
the Markov inequality needs to be applied twice to convert the
average guarantee to individual guarantee which greatly increases
the $d$ level for individual guarantee to $d^{\frac{1}{3}}$ [15].

                       Generally, the wrong interpretation of a $d\leq\epsilon$ key as an $\epsilon$-uniform key solves the following security problems handily which now have to be dealt with anew. As we shall see, either the solution is not known or appears extraordinarily difficult or the resulting situation is very unfavorable for such QKD key.
\begin{enumerate}[(i)]

\item A primary point is what numerical value of $d$ is adequate
for security. The value $d=10^{-20}$ suggested in ref. [19] is
good under the wrong interpretation if only a one-shot trial is
involved, and thus a perfect key is for all practical purposes
guaranteed except for the PAC and $K$ averaging involved. However,
the resulting level of effective $d'\sim 10^{-7}$ for individual
guarantee from Markov inequality is far from adequate. If we take
$10^{-15}$ to be an effective one-shot impossibility, a $d$-level
less than $10^{-40}$ is required. Even such relatively very large
$d$ level (for $n$ exceeding 1,000) cannot be remotely approached
in a concrete QKD protocol. To ensure a "near-uniform" key of
$d\sim2^{-n}$, it follows from (9) that one needs $d\sim
10^{-300}$ for $n=1,000$. The most up-to-date single-photon BB84
protocol analysis (with many invalid steps) already gives zero net
key rate for $K$ at $d=10^{-15}$ [20]. If a near-uniform key is
desired, $d=10^{-20}$ would give no more than 22 bits in
principle, not enough to cover the message authentication key bits
not yet accounted for. The situation is dire for repeated uses of
the QKD system. This issue of actual numerical values will be
further elaborated in section 8.

\item KPA security becomes now a serious issue due to (11), but is
resolved in principle satisfactorily from our (12) for the case of
sequence (subset of $K$) estimation by Eve.

\item As mentioned in section 3, Eve's BER needs to be bounded
even when she fails to identify a subset of $K$ correctly. There
is no such problem for a perfect key, of course. Apart from the
whole $K$ before it is used [15], there is \textit{no} such bound
known in more general situations including KPA.

\item Universal composition which follows immediately from the
wrong interpretation of $d$ [21] is lost. Each application of $K$
has to be examined by itself to see what operational guarantee
would result, in particular under possible quantum information
locking leak.

\item Security proofs involving the error correcting code and the
subsequent privacy amplification code are seriously affected when
$K$ is not perfect. We will discuss this separately in the
following section 5. This appears to be an issue for which
\textit{no} satisfactory solution can be found without the
introduction of major new cryptographic technique. \item An
imperfect key can have a very serious detrimental security effect
on its use in an information theoretically secure message
authentication code (MAC) [18, 22-24], especially for the
relatively large $d$-level that can be obtained in concrete QKD
protocols as discussed in (i) above. A typical MAC with ITS
consists of a keyed hash family with key $K_h$ and often another
key $K_t$ for OTP the authentication tag. The security against
impersonation and substitution attacks is guaranteed through an
$\epsilon$-ASU hash family, in which $\epsilon$, $0<\epsilon<1$,
upper bounds Eve's success probability and $\epsilon$ itself is
lower bounded by $\frac{1}{|t|}$. When $K_h$ has a statistical
distance $\epsilon_h$ from uniform, or equivalently has a
$d$-level $\epsilon_h$ in the quantum case, Eve's success
probability may reach 1 for some tag sequences as in the privacy
case (11) above [18]. Upon tag average similar to (12), an
$\epsilon$-ASU family becomes  an $\epsilon +\epsilon_h$-ASU
family [25]. When $K_t$ with $d=\epsilon_t$ is used, it becomes an
$(\epsilon+m\epsilon_t)$-ASU family when the hash function is used
$m$ times [24]. Thus, a lower limit is now set by $\epsilon_h$ or
$\epsilon_t$ however long the authentication tag is! There is
\textit{no longer} a security parameter for MAC since the QKD
$\epsilon$-key itself has none. Typical tag length of 64 bits
already requires $\epsilon_h$ or $\epsilon_t$ be at the level of
$d\sim 10^{-20}$ for individual guarantee, tens of orders of
magnitude from that derived for just theoretical single-photon
BB84 [20] and fifty orders of magnitude from the current
experimental level. When the tag average us taken into account it
could reach one hundred orders of magnitude. Thus, a QKD key so
generated does not and \textit{cannot} measure up to the key
needed in most common MAC.
\end{enumerate}

\section{ERROR CORRECTION AND PRIVACY AMPLIFICATION PROBLEMS}
                  In this section we will show explicitly, for the first time since the ECC and PAC problems were originally indicated in [7] ten years ago and elaborated somewhat in [25], that these problems appear quantitatively insurmountable. The main culprit is the ECC problem, the PAC problem can by itself be handled as in [14] from a proper EC treatment that appears impossible to carry out. The ad hoc and invalid ECC treatment in the literature thus serves two purposes, to quantify the information leak (or key cost) from ECC and to allow the application of PAC theory, without which there would be no quantification of the security and key rate of a QKD protocol.

                   In the QKD literature the Cascade reconciliation protocol was popular, but it has numerous invalid steps on estimating the information leak to Eve which cannot be usefully bounded [26]. This information leak from error correction is simply neglected in the earlier general security papers. More recently it is given by the ad hoc formula, for $1\leq f\leq 2$,
\begin{equation}
leak_{EC}=f\cdot n\cdot h(Q)
\end{equation}
Here $Q$ is the quantum bit error rate the users measure and $h$
is the binary entropy function. It is ad hoc because the factor
$f$ is arbitrarily taken to represent the effect of a finite
protocol with $n\cdot h(Q)$ itself taken to be the asymptotic
leak. What is the derivation of (13) under a general attack, just
asymptotically? There is none offered in the literature, this
crucial difficulty not mentioned at all! In [20] the whole book
[17] is referred to for $n\cdot h(Q)$ but [17] does not treat such
problem. In particular, the memoryless channel treated in [17]
simply does not apply to joint attack. In [3] there is no formula
given for $leak_{EC}$ and thus the results are true by definition,
except it is then not shown that the final key rate is positive.

                     The discussion in [25] would not be repeated here on exactly what $leak_{EC}$ may be under a general attack. Under collective attack only, it may appear that (13) for $f=1$ may be derived asymptotically by OTP the parity check digits of a linear ECC with uniform key bits. When the key bits are not uniform to Eve as those from a QKD key, there is \textit{no} quantification at all on what the resulting $leak_{EC}$ is, as point (v) of the last section indicates. However, even when uniform bits are available for padding the problem is far more complicated even just for collective attacks, for the following reason.

                     Let $\rho_x$ be the density operator of Eve's probe which depends on the data $x$ chosen by Alice on the sifted key. With a specific ECC, say the $ith$ one among a given set $\mathscr{I}$ of possible ECC, the state becomes $\rho^i_x$, $i\in\mathscr{I}$. Padding the parity digits merely shows that only $p_i$ but not $i$ is known to Eve. Thus Eve's state $\rho_x$ is transformed to $\rho'_x$
\begin{equation}
\rho'_x=\sum_ip_i\rho^i_x
\end{equation}
There may still be information leak from the change of $\rho_x$ to
$\rho'_x$ in addition to the padding bit cost, because $\rho_x^i$
with an ECC may leak a lot more information to Eve than $\rho_x$
itself for at least some or possibly all $i\in\mathscr{I}$. This
happens because the ECC allows Eve to correct her erros too. At
any rate $\rho_x'$ needs to be dealt with for security analysis
after ECC, not $\rho_x$. It appears nothing general can be derived
without attention to specific family $\mathscr{I}$ of the chosen
ECC for dealing with $\rho'_x$.

                                 This problem spills into the PAC one as follows. From [14, 27] one bounds the input state minimum entropy (equivalently $\overline{P_1}$) or its $\epsilon$-smooth generalization, and an output $K$ can be guaranteed with a certain $d$ level from universal hashing. Padding the ECC parity digits, however, does not imply one can use the $\rho_x$ bound for the PAC input. One has to use that on $\rho'_x$ from (14). Incidentally, it is also clear that the ECC output state, which is the PAC input state, has to be bounded appropriately no matter what $leak_{EC}$, whether correct or not, is being used. Since the incorrect input state $\rho_x$ is used for the PAC input, the PAC guarantee itself becomes \textit{invalid}.

                                There are only two approaches to finite protocols that offer a more or less complete treatment of a QKD protocol apart from message authentication. We have discussed the Renner approach above. Hayashi completes and generalizes the Shor-Preskill [28] approach to directly treat BB84 protocols, with different ways of accounting for the ECC [29] and PAC leaks [30]. However, the ECC leak bound cannot be quantitatively carried out and (13) is used instead [31]. It should be clear that (14) or its equivalent in any QKD approach would constitute a major obstacle for quantifying security with ECC, and hence PAC also. They do \textit{not} appear to be amenable to quantitative treatment for the long ECC needed in QKD.

\section{PROBLEMS OF CONTINUOUS VARIABLE QKD}
       There have been much recent work on continuous variable QKD (CV-QKD) [3] due to its immunity to detector blinding attacks [32] from homodyne detection. There is a special $leak_{EC}$ problem in CV-QD but already the issue from (14) cannot be dealt with for whatever reconciliation procedure. There is a further \textit{robustness} problem that has never been addressed but which would render CV-QKD impractical, and it apparently cannot be overcome, as follows.

                       Under an active heterodyne intercept-resend attack, the security analysis of CV-QKD with mutual information or whatever criterion in the literature becomes invalid because Eve has fundamentally altered the channel. She would get the data better than Alice or Bob in either the direct or the reverse reconciliation approach [33]. For security she must be caught during the checking phase of the protocol. However, it is practically impossible to run a protocol with such check for the following reason.

                                  Let $T$ be the system transmittance between Alice and Bob, and $S$ the source photon number. Let $a$ be the source fluctuation or knowledge inaccuracy, $b$ that of $T$, so that the total inaccuracy in the output signal level is $(a+b-ab)ST$. Thus, for 1\% uncertainty in both $S$ and $T$, the output uncertainty is $\sim 2\%$. Whenever $ST$ is significantly less than $\frac{1}{2}$, the users cannot tell Eve's presence with her heterodyne-resend attack, thus establishing a loss limit on security. Whenever $(a+b-ab)ST$ is bigger than $\frac{1}{4}$, the users cannot distinguish Eve's attack from uncertainty in $ST$. This sets a strict upper limit on $S$.

                                  In reality, the users do not know many system parameters to 1\% even when they do not fluctuate or change from bit to bit. Line loss is especially widely uncertain, even in a fiber. We cannot assume it is constant during the execution of a QKD protocol. Such small uncertainty does not matter in ordinary optical communications but matters a lot for QKD. During checking for Eve's presence it may well happen that no threshold can be found with both acceptable security and low enough false alarm rate (leading to protocol being aborted) that renders the protocol not much more inefficient than its already very low efficiency compared to ordinary optical communications. Similar false alarm issue is also present in BB84 though not as seriously. Note that key bits spent in false alarm rounds are cost of the QKD protocol. It  seems that single-photon or low signal level cryptography, QKD or whatever, is a bad engineering idea.

\section{A LIST OF SOME OTHER QKD SECURITY PROOF PROBLEMS}
                             In this section we list some further major security proof problems with brief comments. More detailed discussions will be referenced or provided in the future.
 \begin{enumerate}[(1)]

\item No proof other than occasional invalid brief remarks has
ever been provided to show why channel loss has no security effect
other than throughput reduction, even after reduction of the state
space to include at least the vacuum state. See [34] for some
specific discussion and more will be provided elsewhere.

\item The use of decoy states for multi-photon sources has widely
been assumed [3] to lead to general security for a 0.5 average
photon number Poisson source. Some problems on such claim are
described in [35] and more will be given in the future. Basically,
lots of attacks from Eve have not been accounted for. Furthermore,
there is no known concrete protocol that gives the claimed key
rate and accessible information level even if that exists from
[28, 36]. Privacy amplification with decoy states cannot be
quantified for several reasons, in contrast to the claim in [36],
including one similar to (14) that arises from multi-photon leak
specifically and not just from ECC.

\item The classical inference from checking qubits to the sifted
key qubits underestimates the error significantly. A similar error
is made in [20] in the bounding of $H_{min}$. See [5] for a brief
discussion with further details to be given elsewhere.

\item  The symmetrization argument for general attack bounding is
not applicable to the concrete BB84 protocols. Again see [5] and
future treatment.

\item Eve's quantum probe needs to be considered for any
application of $K$, it cannot be argued away from "universal
composition" without the incorrect $d$ interpretation. The
phenomenon of quantum information locking has to be covered. In
particular, the situation has not been treated for the application
of $K$ to conventional MAC with ITS. The treatments in [18, 21-24]
are classical.

\item Detector blinding attacks [32] show that there is no
security proof without modeling the relevant detector behavior
explicitly, which has not been carried out. There is a more
general problem of model completeness and related issues [37] that
is absent or much less serious in mathematical cryptography and
when larger signals are used in physical cryptography. These
issues cannot be avoided in the "device independent" type
approaches. Indeed, such issues can be traced to the use of small
single-photon signals which are in principle sensitive to minute
disturbance in order that the disturbance-information tradeoff
principle can function.

\item We include here the problems of actual experimental
implementations with quoted theoretical claims about security
level and key rate that are not justified even according to the
theory literature. Thus, the NEC system security claim [38] does
not account for ECC and PAC leaks, and the Toshiba UK system [39]
cannot derive from [28] and [36] their key rate and security level
( and their decoy state generalization) because such results are,
in addition to being invalid as discussed some above, are only
claimed to have been established for CSS codes as ECC and PAC that
are not implemented in their system.
\end{enumerate}
\section{NUMERICAL VALUES AND INDIVIDUAL GUARANTEE}
The claimed theoretical single-photon BB84 security level of up to
$d=10^{-14}$ [20], or even the often claimed practically achieved
level of $d=10^{-9}$ [40] for higher key rates, may seem adequate
under the wrong interpretation of $d$ that $K$ is
$\epsilon$-uniform. Perhaps a probability of $10^{-14}$ is
synonymous with practical impossibility, for a \textit{single}
trial. If such probability level is sufficient, no one would need
a 64 bits or much longer key. It is easily seen that just mere
repetitions would render such probability unacceptable for
security. If 100 rounds per second are carried out in a QKD
system, one day of operation would yield $\sim10^7$ rounds. Thus,
even under the wrong $\epsilon$-uniform interpretation the
security level is not adequate for many applications.

  The situation is worse because there are two separate averages over different random variables in the $d$ of a QKD guarantee $d\leq\epsilon$. One of them is from averaging over PAC that is present in the wrong interpretation. One is averaging over $K$ that is not present in the wrong interpretation, because $K$ is then perfect with a high probability and one can just treat it as uniform for all its values $k$. Even in ordinary manufacturing, individual guarantee is used for quality control. Thus, one needs to convert an average guarantee to an individual one for a QKD key $K$ via Markov inequality, that for a random variable $Z$ one has
\begin{equation}
Pr[|Z|\geq\delta]\leq\frac{E[|Z|]}{\delta}
\end{equation}
  When (15) is applied to minimize the total "failure probability" [15,41], the effective $d$ level becomes $d^{\frac{1}{2}}$ for the wrong interpretation and $d^{\frac{1}{3}}$(but not $d^{\frac{1}{4}}$) under (12). Just from the square root of the wrong interpretation, the effective individual guarantee level as compared to the uniform is unacceptable at $10^{-14}$, not to say from the actual cube root. In effect, there could be many more breaches under an average guarantee and to ensure otherwise, the average level itself has to be further reduced. If one considers $10^{-9}$  as the current state of art (though that is invalidly derived as we discussed in this paper) and $10^{-15}$ for individual guarantee as proper goal, there is a 36 orders of magnitude gap toward the goal. New cryptographic technique or principle is clearly called for to bridge such gap.

  In sum, not only is there a fundamental tradeoff between security level and key rate in QKD protocols of finite $n$, there is such a tradeoff asymptotically also and there is no security parameter which can be independently varied to increase security without affecting key rate. The precise quantitatively behavior is very unfavorable as illustrated above and in point (i) of section 4.

\section{REMEDY VIA KCQ AND DBM}
The different quantum cryptographic approach of KCQ (keyed
communication in quantum noise) [7,9] was originally developed to
alleviate the inefficiency, sensitivity (lack of robustness), and
infrastructure incompatibility (commercially) of BB84 type
protocols to make quantum cryptography practical. The original
version of Alpha-Eta for direct encryption [42] has been
extensively developed, in particular by the US company Nucrypt,
which can be called PSK-Y00 to distinguish it from other signal
set choices based on the same principle such as ISK-Y00 [43], CPPM
or PPM-Y00 [9], and the QAM-Y00 being investigated in Japan. I
would like to separate the term "QKD" for protocols that depend on
the disturbance-information tradeoff principle for security, in
contrast with the term "KCQ" that utilizes quantum effects
associated with the optimal quantum receiver principle for quantum
detection [44] which does not require checking disturbance. KCQ is
nevertheless fully quantum with no classical analog and is capable
of delivering ITS as in QKD. It is fully compatible with ordinary
optical communications through fibers or other media.

  However, great difficulties are encountered in providing general security proofs to KCQ protocols, although under "collective attacks" all protocols, classical or quantum, can be readily proved secure apart from the ECC problem for finite protocols. The DBM (decoy bits method, which has nothing to do with decoy states in QKD) approach is introduced to make general rigorous proof possible. It is a widely applicable technique in both classical and quantum cryptography, and will be presented in detail elsewhere. Here we indicate how BB84 could be modified to become not just the
qb-KCQ protocol in [9] but also by DBM that allows a different
rigorous approach to security without disturbance-information
tradeoff: simply use a pseudo-random number generator with a
shared secret seed key to pick the qubits as sifted key. The
channel characterization may be done separately or from the unused
qubits. A final message authentication check on the generated key
should (always) be employed. Its quantitative ITS will be given in
future papers.

\section{CONCLUSION}
Cryptography is a tricky subject, physical and quantum
cryptography more so from the added essential physical features on
top. The problems involved are far from merely mathematical or
physical. It involves conceptual issues on the relation between
mathematics and the real world in ways not encountered in other
fields and not discussed in books and articles. The numerous
erroneous claims in QKD arise from that in part, the prevalent
wrong interpretation of an $\epsilon$-secure key as an
$\epsilon$-uniform key being a good example. Since cryptography is
a serious matter, we must scrutinize our security arguments and
pay due attention to Eve's viewpoint to ensure our security claim
is validly established. In this paper we have presented some
apparently very serious difficulties in providing valid
quantitative security claims in QKD. The keys given in the
analyzed theoretical protocols cannot be considered secure even if
the derivation is valid, while many steps in the analysis are
actually invalid. Claims that the corresponding experimental
systems are secure in principle are not founded. Hopefully the
reader would be motivated to look seriously into alternative
approaches, not just for security but also for the very relevant
issues of efficiency, robustness, and whether quantum cryptography
does provide a sensible real world alternative in various
applications.

\section*{ACKNOWLEDGMENTS}
I would like to thank Greg Kanter and Aysajan Abindin for useful
discussions. This work was supported by AFOSR and DARPA.

\section*{GLOSSARY}
\begin{longtable}[c]{p{50pt} p{175pt}}
DBM & decoy bits method
\\ECC & error correcting code
\\KCQ & keyed communication in quantum noise
\\KPA & known-plaintext attack
\\MAC & message authentication code
\\OTP & one time pad
\\PACE & privacy amplification code
\\QKD & quantum key distribution
\end{longtable}

\end{document}